\documentclass[12pt]{article}
 \pdfoutput=1
\usepackage[table]{xcolor}
\usepackage{amsmath,amssymb,exscale}
\usepackage{graphicx}
\usepackage{epsfig}
\usepackage{multicol}
\usepackage{color}
\usepackage{mathrsfs}
\usepackage{blindtext}
 \usepackage{fancyhdr}
\usepackage{hyperref}
\usepackage{cite}
\usepackage{mathtools}
\usepackage{amsmath}
\usepackage{rotating,slashed,amsmath,charter,xcolor,catchfilebetweentags,ifluatex}

\usepackage{graphicx}
\usepackage{sidecap}

\usepackage[latin1]{inputenc} 
\usepackage{multicol}  
 
 \usepackage{titlesec}
 
 \usepackage{rotating,slashed,xcolor,amsfonts,expdlist,charter}

\numberwithin{equation}{section}

\usepackage{xcolor}
\usepackage{sectsty}

\usepackage{mdframed}
\usepackage{titletoc}

\definecolor{secnum}{RGB}{13,151,225}
\definecolor{ptcbackground}{RGB}{212,237,252}
\definecolor{ptctitle}{RGB}{0,177,235}

\titlecontents{lsection}
  [5.8em]{\sffamily}
  {\color{secnum}\contentslabel{2.3em}\normalcolor}{}
  {\titlerule*[1000pc]{.}\contentspage\\\hspace*{-5.8em}\vspace*{5pt}%
    \color{white}\rule{\dimexpr\textwidth-15.5pt\relax}{1pt}}

\usepackage{hyperref}
\hypersetup{colorlinks,bookmarksopen,bookmarksnumbered,citecolor=blus,
linkcolor=redy,pdfstartview=FitH,urlcolor=blus}
\usepackage{slashed}

\definecolor{blus}{cmyk}{1,0.9,0,0.1}
\definecolor{verdes}{cmyk}{0.99,0,0.59,0.65}
\definecolor{rossos}{cmyk}{0,1,1,0.55}
\definecolor{redy}{cmyk}{0,1,1,0.7}
\definecolor{greeny}{cmyk}{0.99,0,0.59,0.98}
\definecolor{green-go}{cmyk}{0.79,0,0.59,0.5}

\usepackage{titlesec}

\def\Lag{\mathscr{L}}

\newcommand{\beq}{\begin{equation}}
\newcommand{\eeq}{\end{equation}}

\def\hhref#1{\href{http://arxiv.org/abs/#1}{arXiv:#1}} 

 \def\Lag{\mathscr{L}}
 
\newcommand{\tmtextbf}[1]{{\bfseries{#1}}}
\newcommand{\tmtextrm}[1]{{\rmfamily{#1}}}
\def\bp{\bar M_{\rm Pl}}

\def\be{\begin{equation}}
\def\ee{\end{equation}}
\def\ba{\begin{array} }

\def\bac{\begin{array} {c}}
\def\bacc{\begin{array} {cc}}
\def\baccc{\begin{array} {ccc}}
\def\bacccc{\begin{array} {cccc}}
\def\ea{\end{array}}
\def\bea{\begin{eqnarray}}
\def\eea{\end{eqnarray}}

\definecolor{red}{rgb}{1,0,0}

\def\psl{\hbox{\hbox{${p}$}}\kern-1.9mm{\hbox{${/}$}}}
\def\dsl{\hbox{\hbox{${\partial}$}}\kern-2.2mm{\hbox{${/}$}}}
\def\Dsl{\hbox{\hbox{${D}$}}\kern-2.6mm{\hbox{${/}$}}}

\def\Lag{\mathscr{L}}

\newcommand{\gappeq}{{\rlap{{\raise}.5ex\text{\ensuremath{>}}}{{\lower}.5ex\text{\ensuremath{\sim}}}}}
\newcommand{\lappeq}{{\rlap{{\raise}.5ex\text{\ensuremath{<}}}{{\lower}.5ex\text{\ensuremath{\sim}}}}}
\newcommand{\I}{\tmtextrm{1{\kern}-.24em l}}

\begin{document}
\topmargin -1.0cm
\oddsidemargin 0.9cm
\evensidemargin -0.5cm

{\vspace{-1cm}}
\begin{center}

\vspace{-1cm}


 {\LARGE \tmtextbf{ 
\color{rossos} Hearing Higgs with Gravitational Wave Detectors}} {\vspace{.5cm}}\\

\vspace{1.9cm}

{\large  {\bf Alberto Salvio }
{\em  

\vspace{.4cm}

 Physics Department, University of Rome Tor Vergata, \\ 
via della Ricerca Scientifica, I-00133 Rome, Italy\\

\vspace{0.6cm}

I. N. F. N. -  Rome Tor Vergata,\\
via della Ricerca Scientifica, I-00133 Rome, Italy\\ 

\vspace{0.4cm}


\vspace{0.2cm}

 \vspace{0.5cm}
}

\vspace{0.2cm}

}
\vspace{0.cm}

%
%
 %
%
%

\end{center}

%
%
\noindent --------------------------------------------------------------------------------------------------------------------------------

\begin{center}
{\bf \large Abstract}
\end{center}

\noindent The relic gravitational wave background  due to tensor linear perturbations generated during Higgs inflation is computed. Both the Standard Model and a well-motivated phenomenological completion (that accounts for all the experimentally confirmed evidence of new physics) are considered. We focus  on critical Higgs inflation, which improves on the non-critical version and features an amplification of the tensor fluctuations. The latter property allows us to establish that future space-borne interferometers, such as DECIGO, BBO and ALIA, may  detect the corresponding primordial gravitational waves.
  \vspace{0.4cm}

\noindent --------------------------------------------------------------------------------------------------------------------------------

\vspace{1.1cm}


\vspace{2cm}

Email: alberto.salvio@roma2.infn.it


\newpage

\tableofcontents


\section{Introduction}\label{Introduction}

The observation in 2015 of a gravitational wave (GW) event (named GW150914) from binary black hole mergers~\cite{Abbott:2016blz,TheLIGOScientific:2016wyq1}  has opened the era of GW  astronomy and has naturally reinforced the interest in any source of gravitational radiation.  

From the point of view of particle physics there are several interesting phenomena that can generate GWs within the reach of present or future detectors. Examples are provided by strong first-order phase transitions. The phase transitions in the Standard Model (SM), such as the electroweak (EW) one or the confinement/deconfinement transition in quantum chromodynamics (QCD), are not of this type. Therefore, the observation of a GW event that is unambiguously generated by the EW or QCD transitions would be considered as a clear evidence of new physics. 

Other interesting sources of GWs are the quantum tensor fluctuations that are generated during inflation. A detection of such GW background would provide further evidence for inflation and, remarkably, a direct observational confirmation of the quantum nature of gravity. Furthermore, since inflation is often implemented via quantum scalar fields and typically involves extremely high energies,  detecting this type of GWs would give us valuable information on the microscopic theory governing the fundamental interactions between particles.

The experimental confirmation of the existence of the Higgs boson in 2012~\cite{Aad:2012tfa,Chatrchyan:2012ufa} has further increased the interest in the already very popular Higgs inflation (HI) possibility, where the inflaton is identified with the Higgs field~\cite{Bezrukov:2007ep}. In Refs.~\cite{Bezrukov:2007ep,Bezrukov:2009db,Bezrukov:2009-2,Salvio-inf} a viable HI was achieved by introducing a very large non-minimal coupling between the Higgs and the Ricci scalar. 

It was pointed out~\cite{crit}, however, that such a coupling would break perturbative unitarity at a scale much below the Planck mass.  While this does not  directly exclude the HI of~\cite{Bezrukov:2007ep,Bezrukov:2009db,Bezrukov:2009-2,Salvio-inf} as unitarity can still be satisfied non perturbatively, it does prevent us from computing with current techniques the implications of the theory in some sub-Planckian energy regimes and for some background metrics~\cite{Bezrukov:2010jz}. Furthermore, when the mentioned coupling is very large, it is necessary to tune  the high energy values of some parameters in order to preserve the quantum inflationary predictions: if this is not done large higher derivative terms in the effective action  are generated by quantum corrections, changing the output of the model~\cite{Salvio:2015kka,Ema:2019fdd,Calmet:2016fsr} (see also Ref~\cite{Kannike:2015apa} for a related study).

These problems can be avoided~\cite{Salvio:2017oyf,Salvio:2018rv} by realizing HI very close to the critical surface in the parameter space~\cite{Hamada:2014iga,Bezrukov:2014bra,Hamada:2014wna}, which is located at the border between the absolute stability and the metastability of the EW vacuum~\cite{Buttazzo:2013uya}. Indeed, this version, known as critical Higgs inflation (CHI), features a much smaller value of the non-minimal coupling, solving both the problems mentioned above. At the same time, remarkably, CHI leads to larger quantum tensor fluctuations~\cite{Salvio:2017oyf,Salvio:2018rv,Hamada:2014iga,Bezrukov:2014bra,Hamada:2014wna} and one may hope that the corresponding relic background of GWs could be detected, leading to the exciting possibility that the Higgs manifests itself in GW astronomy.

\vspace{0.1cm}

{\it The purpose of this paper is to compute the GW spectrum (as a function of frequency) associated with the tensor linear perturbations in CHI and investigate whether this gravitational radiation can be detected  with future GW interferometers.}

\vspace{0.1cm}

HI (including the critical version) was originally proposed in the SM, which is very successful in predicting the results of most of particle physics experiments. However,  one has to keep in mind that the SM certainly needs to be extended. Indeed, the observed neutrino oscillations and dark matter are enough to establish the existence of some physics beyond the SM (BSM). Due to the very high energy scales during inflation, it is conceivable that the extra degrees of freedom   could affect both the inflationary observables and the relic background of GWs. For this reason it is relevant to investigate the production of inflationary GWs both in the SM and in a scenario that can account for the experimentally confirmed evidence of BSM physics. We do so in this work.

 Here is an outline of the paper.  In Sec.~\ref{Relic background of gravitational waves from inflation} we provide in a model-independent way the gravitational wave spectral density generated by the inflationary tensor fluctuations without committing ourselves to HI  and without assuming a specific model. We do make, however, some simplifying assumptions regarding the properties of the extra light particles and the reheating phase, which are satisfied in HI and CHI (and in the specific models we consider in the rest of the paper). In Sec.~\ref{HIsec} we give the main properties of CHI that are used in the subsequent sections. A description of future GW detectors that would be able to test CHI is given in Sec.~\ref{Relevant gravitational wave detectors}. The SM case is then analyzed in Sec.~\ref{SMcase}, while Sec.~\ref{anuMSM} is devoted to the corresponding analysis in a well-motivated phenomenological completion of the SM below the Planck scale~\cite{Salvio:2015cja,Salvio:2018rv,Salvio:2021puw}.  The conclusions are given in Sec.~\ref{Conclusions}.

\section{Relic background of gravitational waves from inflation}\label{Relic background of gravitational waves from inflation}

 \vspace{0.cm}
 
GWs produced during inflation as quantum tensor fluctuations form today a relic stochastic background of GWs (see Ref.~\cite{Maggiore:2018sht} for a textbook introduction to this topic). If sufficiently strong this background could be detected in future GW experiments.  Therefore, in this section we provide the key formul\ae~for such background in a way that is as model independent as possible, hoping that this could be relevant not only for the specific models of Secs.~\ref{SMcase} and~\ref{anuMSM}, but also for some future activities.

The main quantity we would like to compute is the following dimensionless function of the frequency $f$ (the spectral density)
\be \Omega_{\rm GW}(f) \equiv \frac{f}{\rho_{\rm cr}}\frac{d\rho_{\rm GW}}{df},\ee 
where $\rho_{\rm cr}\equiv	3H_0^2\bp^2$  is the critical energy density, $\bp$ is the reduced Planck mass, $H_0$ is the present value of the Hubble rate and $\rho_{\rm GW}$ is the energy density carried by the stochastic background.
Indeed, this is the quantity normally constrained in the context of interferometers and pulsar timing array experiments. $\Omega_{\rm GW}(f)$ is related to the power spectrum of tensor fluctuations at the present time ${\cal P}_{t,0}(f)$ by
\be \Omega_{\rm GW}(f) = \frac{\pi^2}{3H_0^2}f^2{\cal P}_{t,0}(f). \label{OmegaE1}\ee
 In turn ${\cal P}_{t,0}(f)$ can be expressed in terms of the power spectrum of tensor fluctuations ${\cal P}_{t,{\rm in}}(f)$ (at the time when the  mode with frequency $f$ re-entered the horizon after inflation) through the transfer function ${\cal T}_{\rm GW}$:
\be {\cal P}_{t,0}(f) = |{\cal T}_{\rm GW}(f)|^2 {\cal P}_{t,{\rm in}}(f).\label{Pt0E1}\ee

Let us recall now the definition of ${\cal T}_{\rm GW}$. As we will write explicitly below, this quantity is related to the tensor fluctuations  of the metric $g_{\mu\nu}$, that appear in\footnote{As usual $\mu,\nu, ...$ are spacetime indices, while $i,j,...$ are space indices. $\delta g_{\mu\nu}$ is the fluctuation of the metric $g_{\mu\nu}$, $\tau$ is the conformal time and the background metric has the spatially-flat Friedmann-Robertson-Walker (FRW) form: 
\be ds^2= a(\tau)^2\left(  d\tau^2  - \delta_{ij}dx^idx^j\right). \label{dsBack}\ee} $\delta g_{ij}(\tau,\vec{x}) = a^2(\tau)h_{ij}(\tau,\vec{x})$. Using the conformal Newtonian gauge, the $h_{ij}$, besides being symmetric ($h_{ij}=h_{ji}$), also satisfy 
 \be h_{ii} =0, \qquad \partial_ih_{ij}=0.\label{Condhij}\ee
 It is convenient to perform a Fourier transform
  \bea h_{ij}(\tau, \vec{x}) =  \int \frac{d^3q}{(2\pi)^{3}}  e^{i\vec{q}\cdot \vec{x}} \sum_{\lambda= \pm 2}  h_\lambda(\tau,\vec{q}) e^\lambda_{ij} (\hat q),  \label{ExpMom}  \eea
where $e^\lambda_{ij} (\hat q)$ are the usual polarization tensors for helicity $\lambda= \pm 2$. We recall that for $\hat q$ along the third axis the polarization tensors that satisfy (\ref{Condhij}) are  given by 
\be e^{+2}_{11} = -e^{+2}_{22} = 1/\sqrt2, \quad e^{+2}_{12} = e^{+2}_{21} = i/\sqrt2, \quad e^{+2}_{3i} =e^{+2}_{i3} = 0, \quad  e^{-2}_{ij} =  (e^{+2}_{ij})^* \label{PolT}\ee 
and for a generic momentum direction $\hat q$ we can obtain $e^\lambda_{ij} (\hat q)$ by applying to (\ref{PolT}) a rotation that connects the third axis with  $\hat q$.
The polarization tensors defined in this way obey 
$ e^\lambda_{ij} (\hat q) (e^{\lambda'}_{ij} (\hat q))^* =2 \delta^{\lambda\lambda'}.  $ The EOM of $h_{\lambda}$  in the vacuum is 
\be h_\lambda''+2{\cal H}h_\lambda' + q^2 h_\lambda' = 0, \label{hlambdaEq} \ee
where ${\cal H}\equiv a'/a$, $q\equiv |\vec{q}|$ and a prime denotes the derivative with respect to $\tau$.
Eq.~(\ref{hlambdaEq}) has to be solved with the initial condition that at the time $\tau_{\rm in}$  the function $ h_\lambda$ reduces to the constant value at  horizon exit after inflation. Both this condition and Eq.~(\ref{hlambdaEq}) depend on  $\vec{q}$ only through $q$ so we can write that $h_\lambda$  is a function of $\tau$ and $q$ only, i.e. $h_\lambda(\tau, q)$. The transfer function is defined by
\be h_\lambda(\tau_0, q) \equiv {\cal T}_{\rm GW}(f) h_\lambda(\tau_{\rm in}, q), \ee
where a subscript $0$ denotes the present time, and $q$ is related to the frequency $f$ via $f\equiv q/(2\pi)$.

The free-streaming of the active neutrinos,  the photon  and other light particles (if any)  could generate a non-vanishing right-hand side of Eq.~(\ref{hlambdaEq}). However, for the frequency range relevant for ground-based  and space-borne interferometers and pulsar timing arrays (i.e. $10^{-9}$\,Hz\,$\lesssim f\lesssim 10^3$\,Hz)
 the neutrino and the photon free-streaming does not affect the amplitude of GWs~\cite{Saikawa:2018rcs}. Should the specific model at hand contain extra light particles, one would have to take into account a possible sizeable effect of these extra species on Eq.~(\ref{hlambdaEq}). 

The GWs that are within the sensitivities of ground-based and space-borne interferometers as well as pulsar timing arrays all correspond to tensor modes that re-entered the horizon during the radiation dominated era much before the time of radiation-matter equality. Solving Eq.~(\ref{hlambdaEq}) for these modes and performing a time average, the transfer function can be approximated by 
\be {\cal T}_{\rm GW}(f)\simeq \frac{a(\tau_q)}{\sqrt2 a_0} =\frac{a(\tau_q)}{\sqrt2} \label{TGWapp}\ee
where $\tau_q$ is the time when the mode with momentum $q$ re-entered the horizon, i.e. ${\cal H}(\tau_q) = q$, and, as usual,  we have conventionally set $a_0=1$.

On the other hand, using entropy conservation and again the fact that the tensor modes re-entered the horizon during the radiation dominated era, one obtains
\be a(\tau_q) \simeq \left(\frac{g_*(T_q)}{\bar g_*}\right)^{1/2}\left(\frac{\bar g^S_*}{g^S_*(T_q)}\right)^{2/3} \frac{H_0 \Omega_R^{1/2}}{q}, \label{atauq} \ee 
where $T_q$ is the value of the (photon) temperature $T$ when the mode with momentum $q$ enters the horizon and $g^{(S)}_*$ is the (entropy) effective  number of relativistic species. These functions of $T$ are defined as usual by
\bea g_*(T)&\equiv& \sum_b g_b \left(\frac{T_b}{T}\right)^4 +\frac78 \sum_f g_f \left(\frac{T_f}{T}\right)^4\\ 
g^S_*(T)&\equiv&  \sum_b g_b \left(\frac{T_b}{T}\right)^3 +\frac78 \sum_f g_f \left(\frac{T_f}{T}\right)^3,
  \eea
 where the sum over $b$ runs over all relativistic bosons, that over $f$ runs over all relativistic fermions (with number of spin or helicity states $g_b$ and $g_f$ and temperatures $T_b$ and $T_f$, respectively).
Moreover, in Eq.~(\ref{atauq}) $\bar g^{(S)}_*$ is the value of $g^{(S)}_*$ at a reference temperature $T_r$ below that of $e^\pm$ annihilation, but such that the three active neutrinos  are still relativistic.  The quantity $\Omega_R$ denotes as usual the ratio of (today's) energy density in radiation to  $\rho_{\rm cr}$, which can be conventionally expressed in terms of the effective number of neutrino species $N_{\rm eff}^{(\nu)}$ and the ratio $\Omega_{\gamma}$ of the photon energy density $\rho_{\gamma}=\pi^2 T_0^4/15$ to  $\rho_{\rm cr}$:
\be h^2\Omega_R  = h^2\Omega_{\gamma} \left[1+N_{\rm eff}^{(\nu)}\frac78 \left(\frac4{11}\right)^{4/3} \right], \label{OmegaRgamma}\ee  
 where as usual $T_0$ is  today's  temperature and  $h\equiv H_0/(100~{\rm km\,s^{-1}\,Mpc^{-1}})$.
 
 Note that $h^2\Omega_R$ can be computed once $N_{\rm eff}^{(\nu)}$ is known because $h^2\Omega_{\gamma}$ is determined by the known value of $T_0$~\cite{Fixsen:2009ug}:
 \be h^2\Omega_{\gamma} = h^2\frac{\rho_{\gamma}}{\rho_{\rm cr}} = \frac{\pi^2T_0^4}{45\bp^2} \left(\frac{h}{H_0}\right)^2,  \ee
 which is independent of the value of $H_0$ because this quantity only appears in the ratio $h/H_0$.  At the temperature $T_r$, by definition, all active neutrinos are relativistic and would contribute to $N_{\rm eff}^{(\nu)}$, but it is important to keep in mind that near the present epoch at least two active neutrinos are non relativistic (see Refs.~\cite{Esteban:2020cvm,deSalas:2020pgw} for recent bounds on their masses).  The value of $\Omega_R$ depends on whether today the lightest neutrino is  relativistic or not.
  In order to compute the active neutrino contribution to $N_{\rm eff}^{(\nu)}$ and to the parameters $\bar g_*$ and $\bar g^{S}_*$, which also enter Eq.~(\ref{atauq}), one should recall that a relativistic neutrino species after $e^\pm$ annihilation features a temperature $T_\nu = (4/11)^{1/3} T$. Moreover, generically $N_{\rm eff}^{(\nu)}$ and the parameters $\bar g_*$ and $\bar g^{S}_*$  can also receive a contribution from extra species that are relativistic at the  temperature $T_0$ and $T_r$, respectively (if any).

In order to use~(\ref{atauq}) we need 
to estimate  the dependence of $T_q$ on $q$. This can be obtained by applying the entropy conservation in the form
\be g_*^S(T_q) T_q^3 a^3(\tau_q) = g_{*0}^S T_0^3. \label{entropyCon}\ee 
The quantity $g_{*0}^S\equiv g_{*}^S(T_0)$ certainly receives a contribution from photons, but can also receive a contribution from one  active neutrino (depending on its mass) and other possible extra species that are relativistic at the present time (if any). By using~(\ref{atauq}) in~(\ref{entropyCon}) one finds
\be T_q = \left(\frac{g_{*0}^S}{\bar g_{*}^S}\right)^{1/3} \frac{c(T_q) T_0 q}{H_0 \Omega_R^{1/2}},  \label{Tq}\ee
where $c(T)$ is a slowly-varying function of $T$ given by
\be c(T)=\left(\frac{g_*(T)}{\bar g_*}\right)^{-1/2}\left(\frac{\bar g^S_*}{g^S_*(T)}\right)^{-1/3}. \label{cTq}\ee
Eqs.~(\ref{Tq}) and~(\ref{cTq}) can be used to estimate $T_q$ as a function of $q$: we can first approximate $c(T_q)$ with its value at the $q$-independent temperature $T_{q0}\equiv T_r$, i.e. $c(T_q)\simeq c(T_{q0})=1$, which gives in the right-hand side of Eq.~(\ref{Tq}) 
\be T_{q1} \equiv \left(\frac{g_{*0}^S}{\bar g_{*}^S}\right)^{1/3} \frac{T_0 q}{H_0 \Omega_R^{1/2}}.\label{Tq1}\ee
Next we approximate $c(T_q) \simeq c(T_{q1})$, which gives through~(\ref{Tq}) 
\be T_{q2} \equiv \left(\frac{g_{*0}^S}{\bar g_{*}^S}\right)^{1/3} \frac{c(T_{q1}) T_0 q}{H_0 \Omega_R^{1/2}}, \label{Tq2}\ee
Reiterating this procedure a few times one obtains a good estimate of $T_q$. In practise, $T_q\simeq T_{q2}$ is already good enough for our purposes.

One can obtain a more convenient expression for $\Omega_{\rm GW}(f)$ by writing 
\be {\cal P}_{t,{\rm in}}(f) = r(q_*)A_{\cal R}(q_*) \left(\frac{f}{f_*}\right)^{n_t}, \label{Ptdec}\ee
where $q_*$ is a pivot scale used in  cosmic microwave background (CMB) observations, $A_{\cal R}$ is the amplitude of the (scalar) curvature perturbation, $r$ is the tensor-to-scalar ratio, $f_* = q_*/(2\pi)$ and $n_t$ is the spectral index of tensor perturbations. In the specific case of single-field inflation (which is the relevant one in CHI)  $n_t = -r/8$, but Eq.~(\ref{Ptdec}) holds even in multi-field inflationary scenarios.
 By inserting~(\ref{Ptdec}) in~(\ref{OmegaE1})-(\ref{Pt0E1}) one obtains
 \be \Omega_{\rm GW} (f) = \frac{\pi^2}{3H_0^2} f^2 |{\cal T}_{\rm GW}(f)|^2 r(q_*)A_{\cal R}(q_*) \left(\frac{f}{f_*}\right)^{n_t}. \label{OGWfin}\ee 
This formula might be changed by the reheating era, but for the frequency range relevant for ground-based and space-borne interferometers as well as pulsar timing arrays, we checked that these corrections are negligible for a reheating temperature $T_{\rm RH}$ above the $10^{10}$\,GeV scale. As we will see in Sec.~\ref{HIsec}, this is the case in HI, which we focus on in this work.
 
 As a final remark for this section, note that $h^2\Omega_{\rm GW}$ as well as $T_q$ and ${\cal T}_{\rm GW}$ are independent of $H_0$ and thus their determination is not affected by the problem of the Hubble tension (see Ref.~\cite{DiValentino:2021izs} for a  recent review).

 \section{(Critical) Higgs inflation}\label{HIsec}
 
 In this work we focus on the case in which the role of the inflaton is played by the Higgs field. However, we would like to provide a model-independent treatment: in this section we do not assume to work within the SM. 
 
 A viable HI means in particular that, in the presence of other scalar fields, the Higgs direction in the potential should be relatively flat compared to the other ones. Moreover, there should be a non-minimal coupling  between the Higgs doublet $H$ and the Ricci scalar\footnote{For the  signature of the metric $g_{\mu\nu}$ we use the mostly minus convention, $g$ is the metric determinant, $R\equiv g^{\mu\nu} R_{\mu\nu}$, where $R_{\mu\nu}\equiv R_{\rho\mu\,\,\, \nu}^{\quad \rho}$, $R_{\mu\nu\,\,\, \sigma}^{\quad \rho} \equiv \partial_{\mu} \Gamma_{\nu \, \sigma}^{\,\rho}- \partial_{\nu} \Gamma_{\mu \, \sigma}^{\,\rho} +  \Gamma_{\mu \, \tau}^{\,\rho}\Gamma_{\nu \,\sigma }^{\,\tau}- \Gamma_{\nu \, \tau}^{\,\rho}\Gamma_{\mu \,\sigma }^{\,\tau}$ and $\Gamma_{\mu \, \nu}^{\,\sigma}$ is the Levi-Civita connection.  }  $R$, which appears in the following term in the  action 
 \be -\int d^4x \sqrt{-g}\xi_H |H|^2 R.\label{xiHterm} \ee
 This term is present together with the standard Einstein-Hilbert term in what is known as the Jordan-frame action.  By performing a field redefinition one can trade the non-minimal coupling with a modification of the other terms (minimally coupled to gravity) where the matter fields appear, such as the scalar  potential. This set of redefined fields is known as the Einstein frame (see Refs.~\cite{Salvio:2018crh,Salvio:2020axm} for a general discussion in the presence of an arbitrary number of scalar, vector and fermion fields).
 As well-known, and as will become clear shortly, the $\xi_H$ term in~(\ref{xiHterm}) is needed to have a sufficiently flat Higgs potential in the Einstein frame.
 
  Also, reheating after inflation occurs very efficiently in HI (both in the critical and non-critical versions)~\cite{Bezrukov:2008ut,BellidoReh}  because the Higgs has sizable couplings to other SM particles. Even restricting to the SM, this leads to a high reheating temperature, $T_{\rm RH}\gtrsim 10^{13}$~GeV. The presence of extra particles in a BSM scenario can further increase $T_{\rm RH}$ because of possible extra Higgs decay channels.

 \subsection{Classical aspects}\label{Classical aspects}

Studying Higgs inflation in the unitary gauge, the Einstein-frame potential of the canonically normalized Higgs field $\phi'$ is given by (see e.g.~\cite{Salvio:2017oyf})
\begin{equation} U_H\equiv \frac{V_H}{\Omega_H^4}=\frac{\lambda_H\phi(\phi')^4}{4(1+\xi_H\phi(\phi')^2/\bp^2)^2},\label{UH}  \end{equation}
where  $\phi$ is the Higgs field non-minimally coupled to gravity that is related to $\phi'$ through 
\begin{equation} \frac{d\phi'}{d\phi}= \Omega_H^{-2} \sqrt{\Omega_H^2 +\frac{3\bp^2}{2}  \left(\frac{d\Omega_H^2}{d\phi}\right)^2 } ,\label{phi'}\end{equation}
$\Omega_H^2$ is defined by 
\begin{equation}  \Omega_H^2\equiv 1+\frac{2\xi_H |H|^2}{\bp^2} \label{transformation}\end{equation}
and $V_H=\lambda_H \phi^4/4$ corresponds to the Higgs potential in the Jordan frame, which involves the Higgs quartic coupling $\lambda_H$.
Eq.~(\ref{UH}) tells us that $U_H$ becomes flat for a sufficiently large $\phi$ (and thus $\phi'$, as this is a monotonically increasing function of $\phi$) at least for $\xi_H>0$.  We impose this positivity condition in the present work.

In a spatially flat FRW geometry the equations for the spatially homogeneous field $\phi'(t)$ and the cosmological scale factor $a(t)$ are  
 \be \ddot\phi' +\frac{\sqrt{3\dot\phi'^2+6U_H}}{\sqrt2\bp}\dot\phi' +\frac{dU_H}{d\phi'}  = 0\label{eq-k=0}\ee
 and 
\be H_I^2=\frac{ \dot\phi'^2+2U_H}{6 \bar M_{\rm Pl}^2},  \label{EE1}\ee
where a dot represents the derivative with respect to cosmic time $t$ and $H_I\equiv \dot a/a$.

In the slow-roll approximation 
the parameters
\be \epsilon_H \equiv\frac{\bp^2}{2} \left(\frac{1}{U_H}\frac{dU_H}{d\phi'}\right)^2, \quad \eta_H \equiv \frac{\bp^2}{U_H} \frac{d^2U_H}{d\phi'^2} \label{epsilon-def}\ee
are  small and one can simplify considerably the calculation of the functions $\phi'(t)$ and $a(t)$ and the inflationary observables.

To be successful, inflation must last long enough, which leads to a lower bound on the number of e-folds
\be N  \equiv \int_{t_b}^{t_e} dt \,  H_I(t), \ee 
where $t_e$ is the time when inflation ends and $t_b$ is the time when the various inflationary observables such as  $A_{\cal R}$, the corresponding spectral index $n_s$ and $r$ are determined through observations. In the slow-roll approximation $N$ is expressed as a function of the field $\phi_b'$ (at $t_b$) rather than as a function of time,

\begin{equation}N=\int_{\phi'_e}^{\phi'_{\rm b}}\frac{U_H}{\bp^2}\left(\frac{dU_H}{d\phi'}\right)^{-1}
d\phi',
\label{e-folds}\end{equation}
where $\phi'_e$ is the field value at the end of inflation.

 \subsection{Quantum aspects and critical version}\label{Quantum aspects and critical version}

At quantum level the formul\ae~of Sec.~\ref{Classical aspects} remain approximately valid except that one must consider $\lambda_H$ and $\xi_H$ as functions of $\phi'$. Since the definition of these functions can be ambiguous~\cite{Bezrukov:2009db,Bezrukov:2009-2, Bezrukov:2014bra,Bezrukov:2014ipa,Bezrukov:2017dyv}
here we adopt the quantization used in~\cite{Salvio:2017oyf,Salvio:2018rv}, which can be embedded in a UV completion of gravity~\cite{Salvio:2014soa,Salvio:2017qkx,Salvio:2019ewf,Salvio:2019wcp} (see Refs.~\cite{Salvio:2018crh,Salvio:2020axm} for reviews). In this approach the $\phi'$-dependence of $\lambda_H$ and $\xi_H$ is obtained by solving the renormalization group equations (RGEs), which depends on the specific model one considers. This procedure, however, applies both to the critical and non-critical version of Higgs inflation.

With this approach $n_s(q_*), A_{\cal R}(q_*)$ and $r(q_*) $  can be computed through 
\begin{equation}n_s(q_*) =1-6\epsilon_H +2\eta_H, \qquad   r(q_*) =16\epsilon_H, \qquad A_{\cal R}(q_*)= \frac{U_H/ \epsilon_H}{24\pi^2 \bp^4},\label{PRsr} \end{equation}
evaluated at $\phi_b'$.

These theoretical predictions must satisfy the observational limits, which put constraints on specific models realizing CHI. In 2018 Planck released its last cosmological observations related to inflation~\cite{Ade:2015lrj}:
\be n_s(q_*)= 0.9649 \pm 0.0042 \, \,  (68\% {\rm CL}), \quad r(q_*)< 0.076 \, \,  (95\% {\rm CL}),\label{nsrobserved}\ee
while for the curvature power spetrum
\be A_{\cal R}(q_*) =(2.10 \pm 0.03) 10^{-9}, \label{PRobserved}\ee 
where the pivot scale  $q_* =0.05~{\rm Mpc}^{-1}$  is used.
    
    The main features of CHI are a bigger value of $r$ and a smaller value of $\xi_H$ compared to the non-critical HI. The former property gives us the hope to detect primordial gravitational waves in the near future, as we will see in the next sections. The latter one pushes the scale of the perturbative breaking of unitarity $\sim \bp/\xi_H$ (pointed out in Refs.~\cite{crit}) very close to the Planck scale, where anyhow new degrees of freedom are needed to UV complete Einstein gravity. As shown in~\cite{Bezrukov:2010jz,Salvio:2018rv} CHI is a viable effective field theory: the cutoff (i.e. the maximal energy below which consistency is possible) is much above the inflationary energy $U_H^{1/4}$. In the rest of the paper we will assume that the new degrees of freedom required by the UV completion do not already appear at or just above $U_H^{1/4}$. The viability of such assumption has been established in~\cite{Bezrukov:2010jz,Salvio:2018rv}. This guarantees that the predictions for the inflationary observables are not significantly altered by the above-mentioned degrees of freedom.

\section{Relevant gravitational wave detectors}\label{Relevant gravitational wave detectors}


How sensitive should a GW detector be in order to  observe the spectral density $h^2\Omega_{\rm GW}$ in~(\ref{OGWfin})? This depends essentially on the four quantities ${\cal T}_{\rm GW}$, $r(q_*)$, $A_{\cal R}(q_*)$ and $n_t$. To make a rough estimate we can consider the SM, which will be analyzed in more detail in Sec.~\ref{SMcase}.
Using~(\ref{TGWapp}) and~(\ref{atauq}), this case corresponds to  ${\cal T}_{\rm GW}\sim 10^{-21}$~Hz$/f$ and inserting in~(\ref{OGWfin}) one obtains the order-of-magnitude estimate
 \be h^2\Omega_{\rm GW} (f) \sim 10^{-16} \left(\frac{r(q_*)}{0.076}\right)\left(\frac{A_{\cal R}(q_*)}{2.10 \times 10^{-9}}\right) \left(\frac{f}{f_*}\right)^{n_t}. \label{OGWest}\ee
 Planck data~\cite{Ade:2015lrj} put some constraints on multi-field inflationary scenarios so we can approximate $n_t\simeq -r/8$, which means that $h^2\Omega_{\rm GW} (f)$ has a weak dependence on $f$ (cf.~the bound on $r$ in~(\ref{nsrobserved})). So using the observational information in~(\ref{nsrobserved}) and~(\ref{PRobserved}), the estimate in~(\ref{OGWest}) tells us that a GW detector should reach a sensitivity corresponding to $h^2\Omega_{\rm GW} \sim 10^{-16}$. This is some orders of magnitude below the sensitivity of ground-based interferometers and  the Laser Interferometer Space Antenna (LISA)~\cite{Audley:2017drz}. However, some future space-borne experiments may reach it.
 
 One of these is the proposed Japanese mission  DECi-hertz Interferometer Gravitational wave Observatory (DECIGO)~\cite{DECIGO}, which would reach the sensitivity  to $h^2\Omega_{\rm GW} \sim 10^{-17}$ around the scale $f\sim 0.1$~Hz. Indeed, the primary objective of DECIGO is to directly observe the beginning of the universe.
 
 Another space-based gravitational-wave detector that can   reach a similar sensitivity is the proposed Big Bang Observer (BBO)~\cite{Crowder:2005nr,BBO2}, which is intended as a follow on mission to LISA. Although LISA's sensitivity is several orders of magnitude away from the required $10^{-16}$, BBO can reach this goal. The Advanced Laser Interferometer Antenna (ALIA)~\cite{Gong:2014mca,Crowder:2005nr} is yet another proposed space-based interferometer with similar characteristics, although with a slightly smaller frequency scale where the maximal sensitivity is reached.

 \section{The Standard Model case}\label{SMcase}
 
 Let us first apply these results to the SM or, more generically, to any model where all extra fields are decoupled. 
 
In this case the temperature-frequency relation in Eq.~(\ref{Tq})
is uniquely determined up to a small uncertainty due to the fact that the lightest neutrino can be relativistic or non relativistic today: 
\be T_{2\pi f}  \simeq 10^7\, \text{GeV} \, \frac{f }{\text{Hz}} \times \left\{ \bac
\hspace{-0.8cm}3.87 \quad \mbox{(relativistic lightest neutrino)}  \\
3.91 \quad \mbox{(non-relativistic lightest neutrino)}  \ea \right.. \label{Tunc}
\ee
Here we  focus on the case $T \gtrsim 100$~GeV as this corresponds to frequencies that space-borne as well (as ground-based) GW detectors are sensitive to.  The uncertainty in~(\ref{Tunc}) can be neglected within the approximations performed in Sec.~\ref{Relic background of gravitational waves from inflation}.
The two cases above, however, lead to a sizeable difference when computing the transfer function because of its dependence on $N_{\rm eff}^{(\nu)}$, cf.~Eqs.~(\ref{TGWapp}),~(\ref{atauq}) and~(\ref{OmegaRgamma}):

\be {\cal T}_{\rm GW} (f) \simeq 10^{-21}\, \frac{\text{Hz}}{f} \times \left\{ \bac
\hspace{-0.8cm}1.25 \quad \mbox{(relativistic lightest neutrino)}  \\
1.13 \quad \mbox{(non-relativistic lightest neutrino)}  \ea \right.. \label{TGWSM}
\ee

The next step is to determine the inflationary observables appearing in~(\ref{OGWfin}). 
We realize CHI in the SM as explained in Ref.~\cite{Salvio:2017oyf}. Regarding the determination of the dependence $\lambda_H$ and $\xi_H$ on $\phi'$, we solve here the RGEs  given in the appendix of~\cite{Allison:2013uaa} with the procedure outlined  in~\cite{Salvio:2017oyf}. Since we are at criticality, the values of the relevant SM couplings at the EW scale are actually chosen differently compared to the current central values, which would correspond to living too far from the border between the absolute stability and the metastability of the EW vacuum. In particular, we set $M_t\simeq 171.04$~GeV, which, following Ref.~\cite{Salvio:2017oyf}, corresponds to the critical point (see Ref.~\cite{Particle data group (top)} for the experimental value of  $M_t$).

\begin{figure}[t]
\begin{center}
 \includegraphics[scale=0.5]{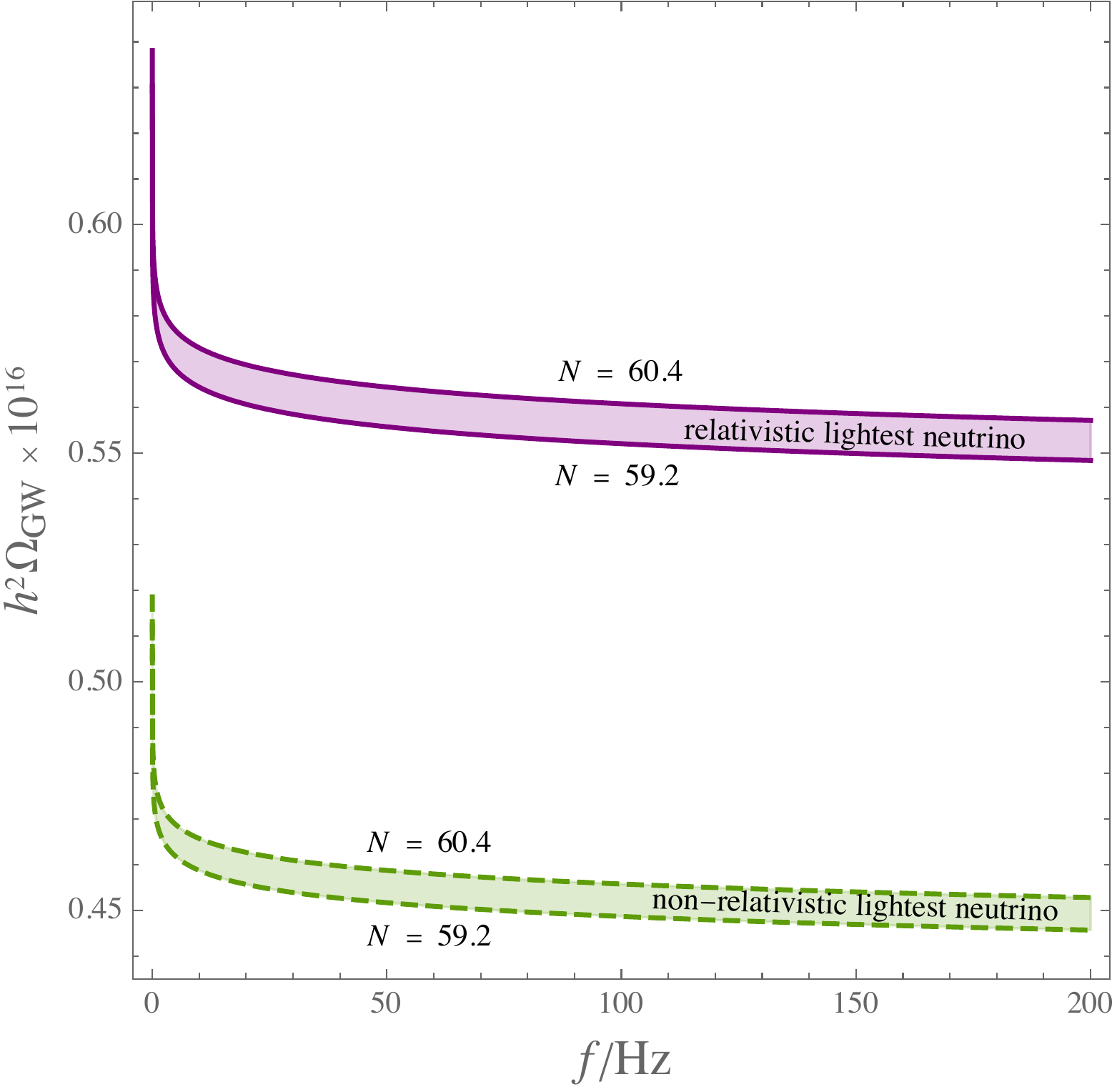}  
 \hspace{0.7cm}  
 \includegraphics[scale=0.287]{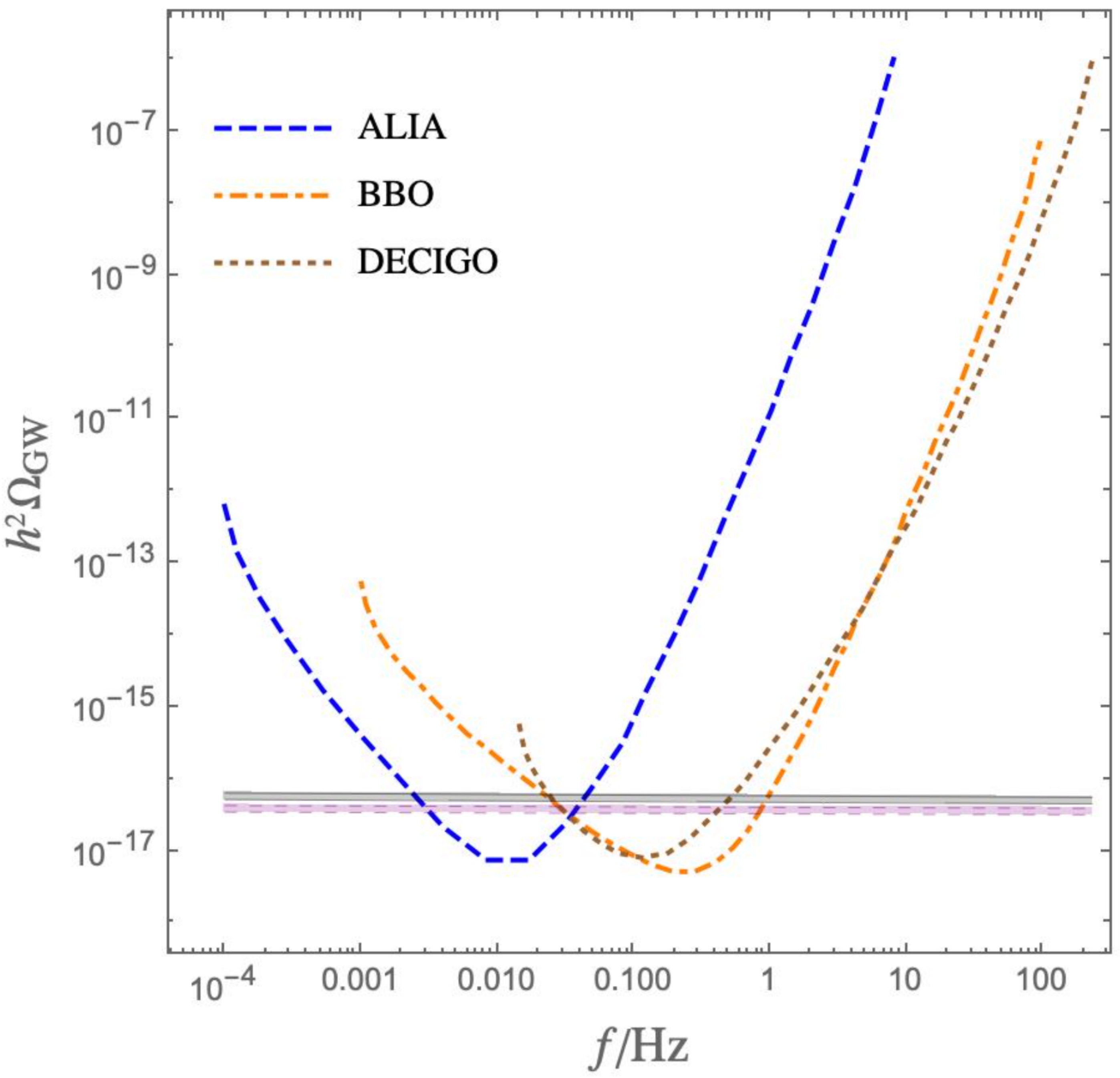} 
 \end{center}

   \caption{\em The GW spectral density as a function of the frequency $f$ in the SM . The two distinct cases of relativistic and non-relativistic lightest neutrino are considered and this corresponds to the width of the quasi-horizontal black band in the right plot (computed with $\xi_H(\bp) \simeq 12.3$). In the left plot we also display the dependence on the number of e-folds $N$ around the value $60$. The SM GW spectral density is also compared to sensitivity curves of future GW detectors in the right plot. In the right plot we also give the predictions in the $a\nu$MSM corresponding to the parameter values quoted in the text (purple quasi-horizontal dashed lines).}
\label{OmegaGW}
\end{figure}

In Fig.~\ref{OmegaGW} we show $h^2\Omega_{\rm GW}(f)$ in two distinct cases: when the lightest (active) neutrino is still relativistic today and when it is not. As shown in the left plot of that figure the result depends on the number of e-folds $N$, which is taken there to be around 60, and, more importantly, on whether today the lightest neutrino is  relativistic or not. In computing $N$ we have taken into account the exact equation of motion of the Higgs field without using the slow-roll approximation. This is significantly more precise than using the slow-roll expression because of the presence of a (quasi) inflection point in the CHI potential~\cite{Salvio:2017oyf}.  In the right plot of Fig.~\ref{OmegaGW} it is shown how the primordial GWs generated by CHI are within the sensitivity of the future space-borne GW detectors BBO, DECIGO and ALIA. The experimental curves are power-law integrated sensitivity curves, which are determined following the method described in~\cite{Thrane:2013oya,Dev:2019njv}. In Fig.~\ref{OmegaGW} we also show the predictions of a BSM scenario that will be discussed in Sec.~\ref{anuMSM}.

Note that  $M_t\simeq 171.04$~GeV that we used in the SM is about $2\sigma$ away from the current central value~\cite{Particle data group (top)}. CHI in the SM also predicts  inflationary observables that are in tension with the most recent Planck bounds~\cite{Ade:2015lrj}, namely~\cite{Salvio:2017oyf}
\be n_s(q_*) \simeq 0.97, \qquad   r(q_*) \simeq  0.076, \qquad A_{\cal R}(q_*)\simeq 2.2\times 10^{-9},\ee
where we used the value of $N$ quoted in Fig.~\ref{OmegaGW}. Similar tensions were also pointed out in Ref.~\cite{Masina:2018ejw}.
However, we should keep in mind that the SM certainly has to be extended because of well-established observations (e.g. neutrino oscillations and DM) as we discussed in the introduction.

 \section{A well-motivated BSM scenario }\label{anuMSM}
 
  We then move to consider the $a\nu$MSM case~\cite{Salvio:2015cja,Salvio:2018rv}. This is an SM extension (defined in Sec.~\ref{Def-anuMSM}) that can account for all the experimentally confirmed signals of BSM physics (neutrino oscillations and dark matter) and can solve other issues of the SM (baryon asymmetry, the strong CP problem\footnote{The other two fine-tuning problems of the SM (the Higgs mass and cosmological constant problems) could be addressed, unlike the strong CP one, with anthropic arguments~\cite{Weinberg:1987dv}.}, the metastability of the EW vacuum and inflation) {\it at the same time}.

 \subsection{Definition of the $a\nu$MSM}\label{Def-anuMSM}

We now give the details  of the $a\nu$MSM that are needed for the following discussion (see Refs.~\cite{Salvio:2015cja,Salvio:2018rv} for further details). To the SM fields one adds three sterile neutrinos $N_i$ and the  fields of the Kim-Shifman-Vainshtein-Zakharov (KSVZ) QCD axion model~\cite{Kim:1979if} (two Weyl quarks $q_1$, $q_2$ neutral under ${\rm SU(2)_{\it L}\times U(1)_{\it Y}}$ and a complex scalar $A$).  This provides a simple but realistic implementation of the Peccei-Quinn mechanism~\cite{Peccei:1977hh}, which solves the strong CP problem.

The SM Lagrangian, $\Lag_{\rm SM}$,  is extended by adding three terms,
 \be \mathscr{L} = \Lag_{\rm SM}+\Lag_{N}+  \Lag_{\rm axion}+ \Lag_{\rm gravity},\label{full-lagrangian} \ee
$\Lag_{N}$ represents the sterile-neutrino-dependent piece:
  \be  i\overline{N}_i \dsl N_i+ \left(\frac12 N_i M_{ij}N_j +  Y_{ij} L_iH N_j + {\rm h.  c.}\right), \ee
where the Majorana mass matrix $M$ is taken diagonal and real, $M=\mbox{diag}(M_1, M_2, M_3),$ without loss of generality.
  $\Lag_{\rm axion}$ is  the KSVZ piece:
  \be \mathscr{L}_{\rm axion} = i\sum_{j=1}^2\overline{q}_j \Dsl \, q_j +|\partial A|^2  -(y q_2A q_1 +h.c.)-\Delta V(H,A),\nonumber \ee
where $\Delta V(H,A)$ is the $A$-dependent term in the classical potential
\be \Delta V(H,A) \equiv \lambda_A(|A|^2-  f_a^2/2)^2 + \lambda_{HA} (|H|^2-v^2)( |A|^2-f_a^2/2),\nonumber \ee
$f_a$ is the axion decay constant and  $v\simeq 174$~GeV is the EW breaking scale. Finally, 
 \be \mathscr{L}_{\rm gravity} = -\left(\frac{\bp^2}{2} +\xi_H (|H|^2-v^2) + \xi_A (|A|^2-f_a^2/2)\right)R -\Lambda,
 \label{gravity-Lag} \ee
where $R$  is the Ricci scalar,
$\xi_H$ and $\xi_A$ represent the non-minimal couplings of $H$ and $A$ to gravity and $\Lambda$ is the cosmological constant. In the $a\nu$MSM the inflaton is identified with the Higgs; like in the SM, it is possible to do so with $\xi_H \sim\mathcal{O}(10)$,  as discussed in Ref.~\cite{Salvio:2018rv}, when we are close to the frontier between the stability and the metastability of the EW vacuum (critical Higgs inflation)\footnote{ A variant of the $a\nu$MSM, where the $M_{ij}$ are generated through the vacuum expectation value of $A$ (which we denote by $\langle A\rangle$) and $|A|$ takes part in a multi-field inflationary dynamics, has been subsequently proposed in~\cite{Ballesteros:2016euj,Ballesteros:2016xej,Ballesteros:2019tvf}. See also Ref.~\cite{Ringwald:2020vei} for a study of the corresponding inflationary GW spectrum. }.

The active-neutrino masses $m_i$ ($i=1,2,3$) are obtained by diagonalizing the matrix \be m_\nu= \frac{m_{D1} m_{D1}^T}{M_1} + \frac{m_{D2} m_{D2}^T}{M_2} + \frac{m_{D3} m_{D3}^T}{M_3}, \label{see-saw}\ee
where the arrays $m_{Di}$ ($i=1,2,3$) are the columns of the Dirac mass matrix
$ m_D = v Y$  i.e. $m_D =\left(\begin{array}{ccc}\hspace{-0.1cm}m_{D1}\,, & \hspace{-0.2cm}m_{D2}\, ,  & \hspace{-0.2cm} m_{D3}\hspace{-0.cm}
\end{array}\right).$
We then express $Y$ in terms of the $M_i$ and  $m_i$  as done in Refs.~\cite{Salvio:2015cja,Salvio:2018rv}.

On the other hand, the PQ symmetry breaking induced by $\langle A\rangle=f_a/\sqrt{2}$ gives to the extra  quarks a Dirac  mass $M_q = y f_a/\sqrt{2}$ and to the extra scalar a squared mass
\be M_A^2 = f_a^2\left(2\lambda_A +\mathcal{O}\left(\frac{v^2}{f_a^2}\right)\right).  \label{MA1} \ee
Since 
\be f_a\gtrsim 10^8~\mbox{GeV} \label{lowerf}\ee 
(see Ref.~\cite{DiLuzio:2020wdo} for a review), the $\mathcal{O}\left(v^2/f_a^2\right)$ term is very small and will be neglected. 

One may wonder whether the PQ symmetry breaking generates a first-order phase transition in this model, such that one may hope to detect the corresponding  GWs as well. The answer to this question is negative~\cite{DelleRose:2019pgi,vonHarling:2019gme} (see also Ref.~\cite{Dev:2019njv} for a previous related discussion), although one can build different BSM scenarios where instead that happens~\cite{DelleRose:2019pgi,vonHarling:2019gme,Ghoshal:2020vud}. Nevertheless, as we show in Sec.~\ref{GWsMSM}, CHI in the $a\nu$MSM can lead to an inflationary background of GWs, which can be observed by future detectors. 
 
 Let us now discuss the other generic observational bounds that are relevant for our purposes (see Refs.~\cite{Salvio:2015cja,Salvio:2018rv} for a  discussion of the remaining observational bounds). 
 
 Regarding the active-neutrinos, here we take  the  currently most precise values reported in~\cite{Esteban:2020cvm,deSalas:2020pgw} for normal ordering (which is currently preferred) of the following quantities: $\Delta m^2_{21}$, $\Delta m^2_{31}$ (where $\Delta m_{ij}^2 \equiv m_i^2-m_j^2$), the active-neutrino mixing angles  and the CP phase in the Pontecorvo-Maki-Nakagawa-Sakata (PMNS) matrix. One must also take into account the cosmological upper bound on the sum of the neutrino masses~\cite{Aghanim:2018eyx}
 \be \sum_i m_i <0.12~\mbox{eV}. \ee 
 
 Regarding the SM sector, we also have to fix the values of the relevant SM couplings at the EW scale, which we identify conventionally with the top mass\footnote{In the $a\nu$MSM (unlike in the SM) we take the central value $M_t\simeq 172.5$ GeV for the top mass.} $M_t\simeq 172.5$ GeV~\cite{Particle data group (top)}. We take the values computed in~\cite{Buttazzo:2013uya}, which expresses these quantities in terms of $M_t$, the Higgs mass $M_h\simeq 125.1$ GeV~\cite{Particle data group}, the strong fine-structure constant renormalized at the $Z$ mass, $\alpha_s(M_Z) \simeq 0.1184$~\cite{Bethke:2012jm} and $M_W\simeq 80.379$ GeV~\cite{Particle data group} (see the quoted references for the uncertainties on these quantities). 
 
 Regarding the axion, besides the lower bound in~(\ref{lowerf}) we have an upper bound on $f_a$  that is obtained by requiring that dark matter is not overproduced. Such bound has a mild logarithmic dependence on $\lambda_A$ (for e.g.~$\lambda_A$ around the scale $10^{-1}$, it is about $5\times 10^{10}$~GeV~\cite{Salvio:2021puw}). When $f_a$ saturates this bound the axion accounts for the whole dark matter~\cite{axionDMmis,axionDMstring}, while for smaller values the rest of dark matter can be explained with the lightest sterile neutrino~\cite{Dodelson:1993je,Shi:1998km,Boyle:2018tzc,Boyle:2018rgh}, forming a two-component DM~\cite{Salvio:2021puw}.
 
Finally, as far as the determination of the dependence $\lambda_H$ and $\xi_H$ on $\phi'$ is concerned, we solve here the RGEs  given in the appendix of~\cite{Salvio:2021puw} as explained in~\cite{Salvio:2018rv}.

\subsection{Inflationary gravitational waves in the $a\nu$MSM}\label{GWsMSM}

  We now apply the formalism of Sec.~\ref{Relic background of gravitational waves from inflation} to determine the primordial GW spectrum generated by CHI in the $a\nu$MSM. 
 
 Here we have an extra very light particle, the axion.  Note that the effect of axion free-streaming is only at the  \% level (see e.g.~Fig.~8 of Ref.~\cite{Ringwald:2020vei}) and can be neglected within the approximations that have been performed here. This means that we can keep solving the vacuum equation (\ref{hlambdaEq}) for the tensor perturbations $h_\lambda$.
  
 A crucial ingredient to compute the GW spectrum of interest is $a(\tau_q)$, which has been estimated in Eq.~(\ref{atauq}). Let us analyse in turn the various ingredients that are needed to determine $a(\tau_q)$ in the $a\nu$MSM.

A first quantity that we need is the 
 reference temperature $T_r$, which enters $a(\tau_q)$ because $\bar g^{(S)}_*$ is the value of $g^{(S)}_*$ at $T_r$. In the  $a\nu$MSM  
we take $T_r$ by definition to be below that of $e^\pm$ annihilation, but such that the three active neutrinos {\it and} the axion are still relativistic. Such a temperature exists thanks to the lower bound on $f_a$  in~(\ref{lowerf}) (which provides an upper bound on the axion mass $m_a$, see again Ref.~\cite{DiLuzio:2020wdo} for a review).

  Moreover, generically $N_{\rm eff}^{(\nu)}$ can  receive an axion contribution, which, however, is very small. The value of $\Omega_R$ that enters $a(\tau_q)$, therefore, essentially only depends on whether today the lightest neutrino is  relativistic or not, like in the SM.

 Other quantities that appear in $a(\tau_q)$ are $\bar g_*$ and $\bar g^{S}_*$; besides the photons and active neutrinos, the QCD axions respecting the lower bound in~(\ref{lowerf}) gives an extra contribution to these parameters (which is also quite small). 
 
   Finally,  in the $a\nu$MSM $g_{*0}^S\equiv g_{*}^S(T_0)$, which is one of the quantities needed to compute the dependence of $T_q$ on $q$, can also receive  a contribution from the axion (depending on its mass).

 These axion contributions can be computed explicitly knowing the decoupling temperature\footnote{A simple formula for $T^a_{\rm dec}$ was proposed by~\cite{Graf:2010tv}
 \be T^a_{\rm dec} \simeq 9.6 \times 10^6\mbox{GeV}\left(\frac{f_a}{10^{10}~\mbox{GeV}}\right)^{2.246}.\ee
 We observe that this formula gives a $T^a_{\rm dec}$ always below  $f_a$ as long as $f_a\lesssim 10^{12}$~GeV, which is required in order not to overproduce dark matter~\cite{Salvio:2021puw}.} $T^a_{\rm dec}$ of axions~\cite{Graf:2010tv,Ringwald:2020vei} and their temperature after decoupling~\cite{DiLuzio:2020wdo}
 \be T_a = T_\nu\left\{\frac{2}{g^{S}_*(T_h)-1}\left[\left(\frac{T}{T_\nu}\right)^3+\frac{21}{8}\right]\right\}^{1/3} =T_\nu \left(\frac{43}{4(g^{S}_*(T_h)-1)}\right)^{1/3}, \ee
where $T_h$ represents some high value of $T$ at a time when all particles shared the same temperature. Thermal axion production has been studied in detail in Ref.~\cite{Salvio:2013iaa} (see also Refs.~\cite{Masso:2002np,Graf:2010tv} for previous studies).

Moreover, realizing CHI in the $a\nu$MSM can also provide values of the inflationary parameters in agreement with the most recent Planck bounds~\cite{Ade:2015lrj}, in addition to accounting for neutrino oscillations, dark matter and baryon asymmetry, solving the strong CP problem and stabilizing the EW vacuum (all at the same time).  This is the case, for example, setting the experimental inputs as in Sec.~\ref{Def-anuMSM} 
and the values of the other parameters as follows: 
$M_1 = 10^{11}\,$GeV, $M_2 = 6.4 \times 10^{13}\,$GeV, $M_3 > \bp$, 
    $f_a \simeq 5 \times 10^{10}\,$GeV, $\lambda_{HA}(M_A)\simeq 0.016$, $\lambda_A(M_A) \simeq 0.10$,  $y(M_A)=0.1$, $\xi_H(M_A)\simeq 14$ and $\xi_A(M_A) \simeq -2.6$. This gives the following values of the inflationary observables
 \be n_s(q_*) \simeq 0.965, \qquad r(q_*) = 0.048, \qquad A_{\cal R}(q_*)=2.1\times 10^{-9} \label{anuMSMinf}\ee
 and a number of e-folds around 60. These values are in perfect agreement with the Planck constraints in~(\ref{nsrobserved}) and~(\ref{PRobserved}).  Moreover, in the $a\nu$MSM all tensions between theoretical predictions and experimental values, which occur in the SM (cf.~the end of Sec.~\ref{SMcase}), are removed.
 The rather high value of $r$ predicted by CHI in the $a\nu$MSM may be tested with other CMB observations, such as those of BICEP and the Keck Array~\cite{BICEP/Keck}. Future planned CMB experiments (e.g.~CMB S4~\cite{Abazajian:2016yjj} and LiteBIRD~\cite{LiteBIRD}) will be sensitive to  smaller values of $r$ and thus will be able to test further CHI, even in a larger class of models.
 
   The corresponding GW predictions are compared with the BBO, DECIGO and ALIA sensitivities in Fig.~\ref{OmegaGW} (right plot). As clear from that plot, these space-borne interferometers have the potential to detect directly these inflationary GWs.

   The model does not predict other GW sources that could hide such signals. Indeed, The QCD and the electroweak transitions are too weak to generate GWs both in the SM (as well known) and in the $a\nu$MSM: the BSM degrees of freedom  are either too massive or they couple too weakly to the SM particles to obtain first order phase transitions, which are necessary to generate observable GWs. Another possible source of GWs is preheating after inflation, which can occur through parametric resonance or tachyonic effects. However, such GWs are always peaked at a much higher frequency, $\gtrsim 10^7$~Hz~\cite{GarciaBellido:2007dg,Figueroa:2017vfa}   and so should not overlap significantly with the signals discussed here.
Finally, the amplitude of gravitational waves produced by QCD axion string-wall networks is too small to be observed even in future gravitational wave detectors with high sensitivities~\cite{Hiramatsu:2012sc}.

\section{Conclusions}\label{Conclusions}

In this work we have explored the possibility to test the Higgs sector through GW interferometers. In particular, we have computed the inflationary GW background due to the tensor linear perturbations coming from quantum fluctuations during HI. We have focused  on CHI because in this case one can push the scale of the perturbative breaking of unitarity due to $\xi_H$ very close to the Planck scale, where anyhow new degrees of freedom are needed to UV complete Einstein gravity. Furthermore, in the CHI, unlike in the non-critical variant, it is not necessary to fine tune  the high energy values of any parameters in order to preserve the quantum inflationary predictions, as commented in the introduction. This is possible because CHI features a moderate value of $\xi_H$, of order 10.

After a rather general calculation of the inflationary GW spectral density $\Omega_{\rm GW}(f)$ due to the tensor linear perturbations, both the SM and a well-motivated phenomenological completion below the Planck scale have been studied in detail. The latter model, dubbed the $a\nu$MSM, features  three right-handed neutrinos and a KSVZ axion sector in addition to the SM fields. Despite being very simple, the $a\nu$MSM  accounts for neutrino oscillations, dark matter, the baryon asymmetry in the universe, solves the strong CP problem  and stabilizes the EW vacuum at the same time. While in the SM CHI is in tension with some particle physics data (in particular $M_t$) and cosmological CMB observations, the $a\nu$MSM  allows us to implement CHI without any tension with our current experimental knowledge.

We have established that, both in the SM and in the above-mentioned BSM scenario, CHI generates an inflationary GW spectrum due to the tensor linear perturbations that is within the reach of some space-borne GW interferometers, such as DECIGO, BBO and ALIA.



\vspace{1cm}

 \footnotesize
\begin{multicols}{2}

\end{multicols}

\end{document}